# Review of New Concepts, Ideas and Innovations in Space Towers

**Mark Krinker**

Member of Advisory Board of Department of Electrical Engineering,
City College of Technology, CUNY, New York, mkrinker@aol.com

## Abstract

Under Space Tower the author understands structures having height from 100 km to the geosynchronous orbit and supported by Earth's surface. The classical Space Elevator is not included in space towers. That has three main identifiers which distingue from Space Tower: Space Elevator has part over Geosynchronous Orbit (GSO) and all installation supported only the Earth's centrifugal force, immobile cable connected to Earth's surface, no pressure on Earth's surface.

A lot of new concepts, ideas and innovation in space towers were offered, developed and researched in last years especially after 2000. For example: optimal solid space towers, inflatable space towers (include optimal space tower), circle and centrifugal space towers, kinetic space towers, electrostatic space towers, electromagnetic space towers, and so on.

Given review shortly summarizes there researches and gives a brief description them, note some their main advantages, shortcomings, defects and limitations.

---------------

## Introduction

**Brief History.** The idea of building a tower high above the Earth into the heavens is very old [1]. The writings of Moses, about 1450 BC, in Genesis, Chapter 11, refer to an early civilization that in about 2100 BC tried to build a tower to heaven out of brick and tar. This construction was called the Tower of Babel, and was reported to be located in Babylon in ancient Mesopotamia. Later in chapter 28, about 1900 BC, Jacob had a dream about a staircase or ladder built to heaven. This construction was called Jacob's Ladder. More contemporary writings on the subject date back to K.E. Tsiolkovski in his manuscript "Speculation about Earth and Sky and on Vesta," published in 1895 [2-3]. Idea of Space Elevator was suggested and developed Russian scientist Yuri Artsutanov and was published in the Sunday supplement of newspaper "Komsomolskaya Pravda" in 1960 [4].  This idea inspired Sir Arthur Clarke  to write his novel, The Fountains of Paradise, about  a Space Elevator located on a fictionalized Sri Lanka, which brought the concept to the attention of the entire world [5].

Today, the world's tallest construction is a television transmitting tower near Fargo, North Dakota, USA. It stands 629 m high and was build in 1963 for KTHI-TV. The CNN Tower in Toronto, Ontario, Canada is the world's tallest building. It is 553 m in height, was build  from 1973 to 1975, and has the world's highest observation desk at 447 m. The tower structure is concrete up to the observation deck level. Above is a steel structure supporting radio, television, and communication antennas. The total weight of the tower is 3,000,000 tons.

At present time (2009) the highest structure is Burj Dubai (UAE) having pinnacle height 822 m, built in 2009 and used for office, hotel, residential.

The Ostankin Tower in Moscow is 540 m in height and has an observation desk at 370 m. The world's tallest office building is the Petronas Towers in Kuala Lumpur, Malasia (2006). The twin towers are 452 m in height. They are 10 m taller than the Sears Tower in Chicago, Illinois, USA.

Current materials make it possible even today to construct towers many kilometers in height. However, conventional towers are very expensive, costing tens of billions of dollars. When considering how high a tower can be built, it is important to remember that it can be built to many kilometers of height if the base is large enough.

**The tower applications.** The high towers (3–100 km) have numerous applications for government and commercial purposes:

•   Communication boost: A tower tens of kilometers in height near metropolitan areas could provide much higher signal strength than orbital satellites.

•   Low Earth orbit (LEO) communication satellite replacement: Approximately six to ten 100-km-tall towers could provide the coverage of a LEO satellite constellation with higher power, permanence, and easy upgrade capabilities.

•    Entertainment and observation desk for tourists. Tourists could see over a huge area, including the darkness of space and the curvature of the Earth's horizon.

•   Drop tower: tourists could experience several minutes of free-fall time. The drop tower could provide a facility for experiments.

•   A permanent observatory on a tall tower would be competitive with airborne and orbital platforms for Earth and space observations.

•   Solar power receivers: Receivers located on tall towers for future space solar power systems would permit use of higher frequency, wireless, power transmission systems (e.g. lasers).

# Main types of Space towers
## 1. Solid towers [6]-[8].

The review of conventional solid high altitude and space towers is in [1]. The first solid space tower was offered in [2-3].The optimal solid towers are detail researched in series works presented in [6-8]. Works contain computation the optimal (minimum weight) sold space towers up 40,000 km. Particularly, authors considered solid space tower having the rods filled by light gas as hydrogen or helium. It is shown the solid space tower from conventional material (steel, plastic) can be built up 100-200 km. The GEO tower requests the diamond.

The computation of the optimal solid space towers presented in [6-8] give the following results:

**Project 1. Steel tower 100 km height**. The optimal steel tower (mast) having the height 100 km, safety pressure stress $K = 0.02$  (158 kg/mm$^2$)($K$ is ratio pressure stress to density of material divided by $10^7$)  must have the bottom cross-section area approximately in 100 times more then top cross-section area and weight is 135 times more then top load. For example, if full top load equals 100 tons (30 tons support extension cable + 70 tons useful load), the total weight of main columns 100 km tower-mast (without extension cable) will be 13,500 tons. It is less that a weight of current sky-scrapers (compare with 3,000,000 tons of Toronto tower having the 553 m height). In reality if the safety stress coefficient $K = 0.015$, the relative cross-section area and weight will sometimes be more but it is a possibility of current building technology.

**Project 2. GEO 37,000 km Space Tower (Mast).** The research shows the building of the geosynchronous tower-mast (include the optimal tower-mast) is very difficult. For $K = 0.3$ (it is over the top limit margin of safety for quartz, corundum) the tower mass is ten millions of times more than load, the extensions must be made from nanotubes and they weakly help. The problems of stability and flexibility then appear. The situation is strongly improved if tower-mast built from diamonds (relative tower mass decreases up 100). But it is not known when we will receive the cheap artificial diamond in unlimited amount and can create from it building units.

**Note: Using the compressive rods** [8]. The rod compressed by gas can keep more compressive force because internal gas makes a tensile stress in a rod material. That longitudinal stress cannot be more then a half safety tensile stress of road material because the compressed gas creates also a tensile radial rod force (stress) which is two times more than longitudinal tensile stress. As the result the rod

material has a complex stress (compression in a longitudinal direction and a tensile in the radial direction). Assume these stress is independent. The gas has a weight which must be added to total steel weight. Safety pressure for steel and duralumin from the internal gas increases $K$ in 35 - 45%.

Unfortunately, the gas support depends on temperature. That means the mast can loss this support at night. Moreover, the construction will contain the thousands of rods and some of them may be not enough leakproof or lose the gas during of a design lifetime. I think it is a danger to use the gas pressure rods in space tower.

## 2. Inflatable tower [9]-[12].

The optimal (minimum weight of cover) inflatable towers were researched and computed in [9-12].

The proposed inflatable towers are cheaper by factors of hundreds. They can be built on the Earth's surface and their height can be increased as necessary. Their base is not large. The main innovations in this project are the application of helium, hydrogen, or warm air for filling inflatable structures at high altitude and the solution of a safety and stability problem for tall (thin) inflatable columns, and utilization of new artificial materials, as artificial fiber, whisker and nanotubes.

The results of computation for optimal inflatable space towers taken from [11] are below.

**Project 1. Inflatable 3 km tower-mast**. (Base radius 5 m, 15 ft, $K = 0.1$). This inexpensive project provides experience in design and construction of a tall inflatable tower, and of its stability. The project also provides funds from tourism, radio and television. The inflatable tower has a height of 3 km (10,000 ft). Tourists will not need a special suit or breathing device at this altitude. They can enjoy an Earth panorama of a radius of up 200 km. The bravest of them could experience 20 seconds of free-fall time followed by 2g overload.

*Results of computations*. Assume the additional air pressure is 0.1 atm, air temperature is 288 $^{\circ}$K (15 $^{\circ}$C, 60 $^{\circ}$F), base radius of tower is 5 m, $K = 0.1$. If the tower cone is optimal, the tower top radius must be 4.55 m. The maximum useful tower top lift is 46 tons. The cover thickness is 0.087 mm at the base and 0.057 mm at the top. The outer cover mass is only 11.5 tons.

If we add light internal partitions, the total cover weight will be about 16 − 18 tons (compared to 3 million tons for the 553 m tower in Toronto). Maximum safe bending moment versus altitude ranges from 390 ton×meter (at the base) to 210 ton×meter at the tower top.

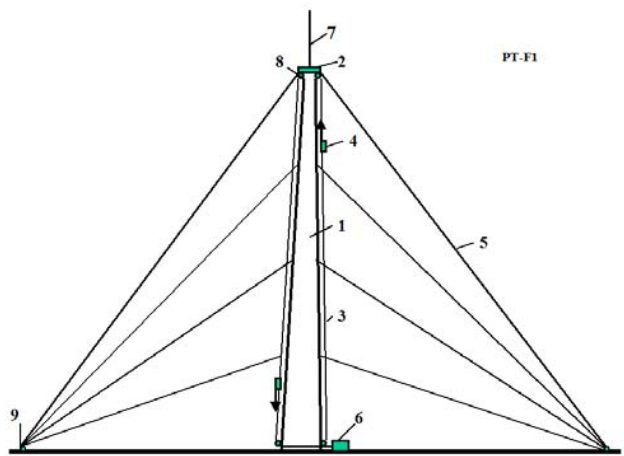

**Fig. 1**. Inflatable tower.
*Notations*: 1 - Inflatable column, 2 - observation desk, 3 - load cable elevator, 4 - passenger cabin, 5 - expansion, 6 - engine, 7 - radio and TV antenna, 8 - rollers of cable transport system, 9 - control.

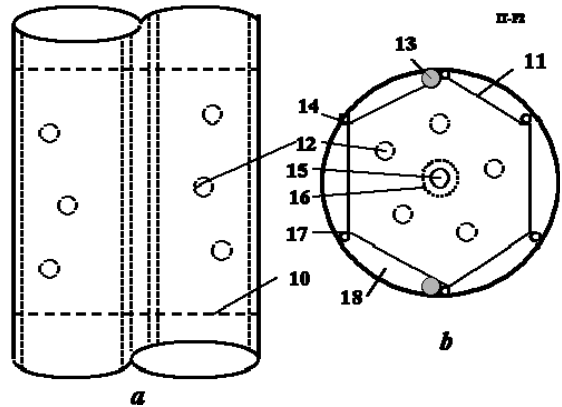

**Fig.2.** Section of inflatable tower. *Notations:* 10 – horizontal film partitions; 11 – light second film (internal cover); 12 – air balls-- special devices like floating balloons to track leaks (will migrate to leak site and will temporarily seal a hole); 13 – entrance line of compression air and pressure control; 14 – exit line of air and control; 15 – control laser beam; 16 – sensors of laser beam location; 17 – control cables and devices; 18 – section volume.

**Project 2. Helium tower 30 km** (Base radius is 5 m, 15 ft, $K = 0.1$)

*Results of computation*. Let us take the additional pressure over atmospheric pressure as 0.1 atm. For $K = 0.1$ the radius is 2 m at an altitude of 30 km. For $K = 0.1$ useful lift force is about 75 tons at an altitude of 30 km, thus it is a factor of two times greater than the 3 km air tower. It is not surprising, because the helium is lighter than air and it provides a lift force. The cover thickness changes from 0.08 mm (at the base) to 0.42 mm at an altitude of 9 km and decreases to 0.2 mm at 30 km. The outer cover mass is about 370 tons. Required helium mass is 190 tons.

**Project 3. Air-hydrogen tower 100 km.** (Base radius of air part is 25 m, the hydrogen part has base radius 5 m). This tower is in two parts. The lower part (0–15 km) is filled with air. The top part (15–100 km) is filled with hydrogen. It makes this tower safer, because the low atmospheric pressure at high altitude decreases the probability of fire. Both parts may be used for tourists.

*Air part, 0–15 km.* The base radius is 25 m, the additional pressure is 0.1 atm, average temperature is 240 °K, and the stress coefficient $K = 0.1$. Change of radius is 25 ÷16 m, the useful tower lift force is 90 tons, and the tower outer tower cover thickness is 0.43 ÷ 0.03 mm; maximum safe bending moment is $(0.5 ÷ 0.03) \times 10^4$ ton×meter; the cover mass is 570 tons. This tower can be used for tourism and as an astronomy observatory. For $K = 0.1$, the lower (0÷15 km) part of the project requires 570 tons of outer cover and provides 90 tons of useful top lift force.

*Hydrogen part, 15–100 km.* This part has base radius 5 m, additional gas pressure 0.1 atm, and requires a stronger cover, with $K = 0.2$.

The results of computation are presented in the following figures: the tower radius versus altitude is 5 ÷ 1.4 m; the tower thickness is 0.06 ÷ 0.013 mm; the cover mass is 112 tons; the lift force is 5 ton; hydrogen mass is 40 tons.

The useful top tower load can be about 5 tons, maximum, for $K = 0.2$. The cover mass is 112 tons, the hydrogen lift force is 37 tons. The top tower will press on the lower part with a force of only 112 – 37 + 5 = 80 tons. The lower part can support 90 tons.

The proposed projects use the optimal change of radius, but designers must find the optimal combination of the air and gas parts and gas pressure.

# 3. Circle (centrifugal) Space Towers [16 - 17]

**Description of Circle (centrifugal) Tower (Space Keeper).**

The installation includes (Fig.3): a closed-loop cable made from light, strong material (such as artificial fibers, whiskers, filaments, nanotubes, composite material) and a main engine, which rotates the cable at a fast speed in a vertical plane. The centrifugal force makes the closed-loop cable a circle. The cable circle is supported by two pairs (or more) of guide cables, which connect at one end to the cable circle by a sliding connection and at the other end to the planet's surface. The installation has a transport (delivery) system comprising the closed-loop load cables (chains), two end rollers at the top and bottom that can have medium rollers, a load engine and a load. The top end of the transport system is connected to the cable circle by a sliding connection; the lower end is connected to a load motor. The load is connected to the load cable by a sliding control connection.

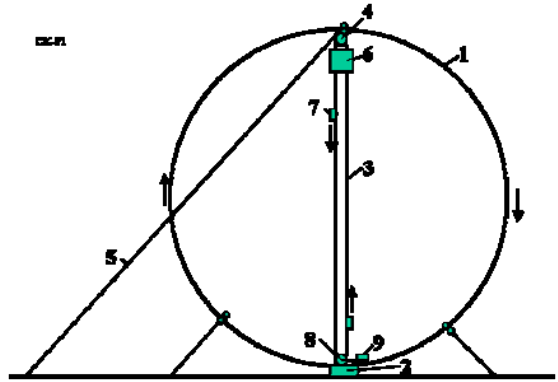

**Fig.3**. Circle launcher (space station keeper) and space transport system. *Notations*: 1 – cable circle, 2 – main engine, 3 – transport system, 4 – top roller, 5 – additional cable, 6 – the load (space station), 7 – mobile cabin, 8 – lower roller, 9 – engine of the transport system.

The installation can have the additional cables to increase the stability of the main circle, and the transport system can have an additional cable in case the load cable is damaged.

The installation works in the following way. The main engine rotates the cable circle in the vertical plane at a sufficiently high speed so the centrifugal force becomes large enough to it lifts the cable and transport system. After this, the transport system lifts the space station into space.

The first modification of the installation is shown in Fig. 4. There are two main rollers 20, 21. These rollers change the direction of the cable by 90 degrees so that the cable travels along the diameter of the circle, thus creating the form of a semi-circle. It can also have two engines. The other parts are same.

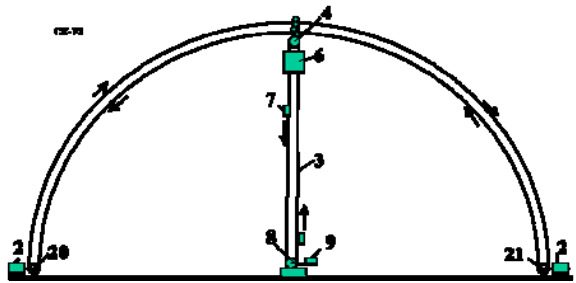

**Fig. 4.** Semi-circle launcher (space station keeper) and transport system. Notation is the same with Fig. 3.1 with the additional 20 and 21 – rollers. The semi-circles are same.

**Project 1. Space Station for Tourists or a Scientific Laboratory at an Altitude of 140 km (Figs.4).**The closed-loop cable is a **semi-circle**. The radius of the circle is 150 km. The space station is a cabin with a weight of 4 tons (9000 lb) at an altitude of 150 km (94 miles). This altitude is 140 km under load.

The results of computations for three versions (different cable strengths) of this project are in Table 1.

**Table 1**. Results of computation of Project 1.

| Variant | $\sigma$, kg/mm$^2$ | $\gamma$, kg/m$^3$ | $K = \sigma/\gamma /10^7$ | $V_{max}$, km/s | $H_{max}$, km | $S$, mm$^2$ |
|---|---|---|---|---|---|---|
| 1 | 2 | 3 | 4 | 5 | 6 | 7 |
| 1 | 8300 | 1800 | 4.6 | 6.8 | 2945 | 1 |
| 2 | 7000 | 3500 | 2.0 | 4.47 | 1300 | 1 |
| 3 | 500 | 1800 | 0.28 | 1.67 | 180 | 100 |

| $P_{max}$[tons] | $G$, kg | Lift force, kg/m | Loc. Load, kg | $L$, km | $\alpha^0$ | $\Delta H$, km |
|---|---|---|---|---|---|---|
| 8 | 9 | 10 | 11 | 12 | 13 | 14 |
| 30 | 1696 | 0.0634 | 4000 | 63 | 13.9 | 5.0 |
| 12.5 | 3282 | 0.0265 | 4000 | 151 | 16.6 | 7.2 |
| 30.4 | 170x10$^3$ | 0.0645 | 4000 | 62 | 4.6 | 0.83 |

| Cable Thrust $T_{max}$, kg, | Cable drag $H = 0$ km, kg | Cable drag $H = 4$ km, kg | Power MW $H = 0$ km | PowerMW $H = 4$ km | Max.Tourists men/day |
|---|---|---|---|---|---|
| 15 | 16 | 17 | 18 | 19 | 20 |
| 8300 | 2150 | 1500 | 146 | 102 | 800 |
| 7000 | 1700 | 1100 | 76 | 49 | 400 |
| 50000 | 7000 | 5000 | 117 | 83.5 | 800 |

The column numbers are: 1) the number of the variant; 2) the permitted maximum tensile strength [kg/mm$^2$]; 3) the cable density [kg/m$^3$]; 4) the ratio $K = \sigma/\gamma \, 10^{-7}$; 5) the maximum cable speed [km/s] for a given tensile strength; 6) the maximum altitude [km] for a given tensile strength; 7) the cross-sectional area of the cable [mm$^2$]; 8) the maximum lift force of one semi-circle [ton]; 9) the weight of the two semi-circle cable [kg]; 10) the lift force of one meter of cable [kg/m]; 11) the local load (4 tons or 8889 lb); 12) the length of the cable required to support the given (4 tons) load [km]; 13) the cable angle to the horizon near the local load [degrees]; 14) the change of altitude near the local load; 15) the maximum cable thrust [kg]; 16) the air drag on one semi-circle cable if the driving (motor) station is located on the ground (at altitude $H = 0$) for a half turbulent boundary layer; 17) the air drag of the cable if the drive station is located on a mountain at $H = 4$ km; 18) the power of the drive stations [MW] (two semi-circles) if located at $H = 0$; 19) the power of the drive stations [MW] if located at $H = 4$ km; 20) the number of tourists (tourist capacity) per day (0.35 hour in station) for double semi-circles.

*Discussion of Project 1.*

1) The first variant has a cable diameter of 1.13 mm (0.045 inches) and a general cable weight of 1696 kg (3658 lb). It needs a power (engine) station to provide from 102 to a maximum of 146 MW (the maximum amount is needed for additional research).

2) The second variant needs the engine power from 49 to 76 MW.

3) The third variant uses a cable with tensile strength near that of current fibers. The cable has a diameter of 11.3 mm (0.45 inches) and a weight of 170 tons. It needs an engine to provide from 83.5 to 117 MW.

The systems may be used for launching (up to 1 ton daily) satellites and interplanetary probes. The installation may be used as a relay station for TV, radio, and telephones.

# 4. Kinetic and Cable Space Tower [13-15].

The installation includes (see notations in Fig.5): a strong closed-loop cable, rollers, any conventional engine, a space station (top platform), a load elevator, and support stabilization cables (expansions). The installation works in the following way. The engine rotates the bottom roller and permanently moves the closed-loop cable at high speed. The cable reaches a top roller at high altitude, turns back and moves to the bottom roller. When the cable turns back it creates a reflected (centrifugal) force. This force can easily be calculated using centrifugal theory, or as reflected mass using a reflection (momentum) theory. The force keeps the space station suspended at the top roller; and the cable (or special cabin) allows the delivery of a load to the space station. The station has a parachute that saves people if the cable or engine fails.

The theory shows, that current widely produced artificial fibers allow the cable to reach altitudes up to 100 km (see Projects 1 and 2 in [14]). If more altitude is required a multi-stage tower must be used (see Project 3 in [14]). If a very high altitude is needed (geosynchronous orbit or more), a very strong cable made from nanotubes must be used (see Project 4 in [14]).

The tower may be used for a horizon launch of the space apparatus. The vertical kinetic towers support horizontal closed-loop cables rotated by the vertical cables. The space apparatus is lifted by the vertical cable, connected to horizontal cable and accelerated to the required velocity.

The closed-loop cable can have variable length. This allows the system to start from zero altitude, and gives its workers/users the ability to increase the station altitude to a required value, and to spool the cable for repair. The innovation device for this action is shown in Fig. 8-6 [14]. The spool can reel up and unreel in the left and right branches of the cable at different speeds and can alter the length of the cable.

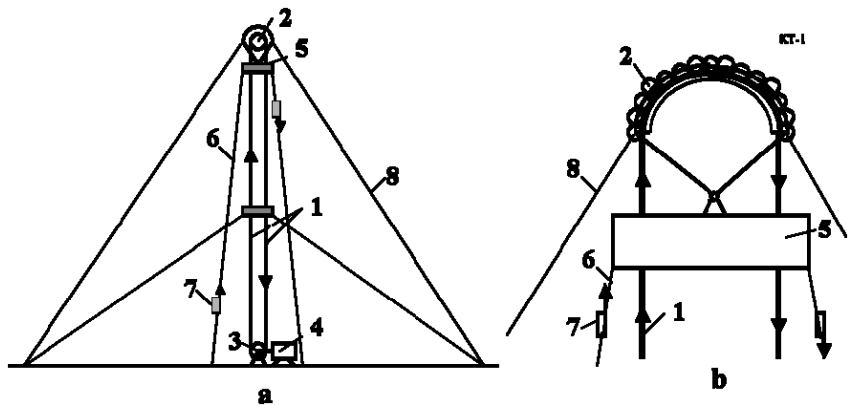

**Fig.5. .a**. Offered kinetic tower: 1 – mobile closed loop cable, 2 – top roller of the tower, 3 – bottom roller of the tower, 4 – engine, 5 – space station, 6 – elevator, 7 – load cabin, 8 – tensile element (stabilizing rope). **b**. Design of top roller.

The safety speed of the cable spool is same with the safety speed of cable because the spool operates as a free roller. The conventional rollers made from the composite cable material have same safety speed with cable. The suggested spool is an innovation because it is made only from cable (no core) and it allows reeling up and unreeling simultaneously with different speed. That is necessary for change the tower altitude.

The small drive rollers press the cable to main (large) drive roller, provide a high friction force between the cable and the drive rollers and pull (rotate) the cable loop.

**Project 1**. **Kinetic Tower of Height 4 km**. For this project is taken a conventional artificial fiber widely produced by industry with the following cable performances: safety stress is $\sigma$ = 180 kg/mm$^2$ (maximum $\sigma$ = 600 kg/mm$^2$, safety coefficient $n$ = 600/180 = 3.33), density is $\gamma$ = 1800 kg/m$^3$, cable diameter $d$ = 10 mm.

The special stress is $k = \sigma/\gamma$ = 10$^6$ m$^2$/s$^2$ ($K = k/10^7$ = 0.1), safe cable speed is $V = k^{0.5}$ = 1000 m/s, the cable cross-section area is $S = \pi d^2/4$ = 78.5 mm$^2$, useful lift force is $F = 2S\gamma(k\text{-}gH)$ = 27.13 tons. Requested engine power is $P$ = 16 MW (Eq. (10), [15]), cable mass is $M = 2S\gamma H = 2\cdot78.5\cdot10^{-6}$ $\cdot1800\cdot4000$ = 1130 kg.

## 5. Electrostatic Space Tower [18]-[19].

**1. Description of Electrostatic Tower**. The offered electrostatic space tower (or mast, or space elevator) is shown in fig.6. That is inflatable cylinder (tube) from strong thin dielectric film having variable radius. The film has inside the sectional thin conductive layer 9. Each section is connected with issue of control electric voltage. In inside the tube there is the electron gas from free electrons. The electron gas is separated by in sections by a thin partition 11. The layer 9 has a positive charge equals a summary negative charge of the inside electrons. The tube (mast) can have the length (height) up Geosynchronous Earth Orbit (GEO, about 36,000 km) or up 120,000 km (and more) as in project (see below). The very high tower allows to launch free (without spend energy in launch stage) the interplanetary space ships. The offered optimal tower is design so that the electron gas in any cross-section area compensates the tube weight and tube does not have compressing longitudinal force from weight. More over the tower has tensile longitudinal (lift) force which allows the tower has a vertical position. When the tower has height more GEO the additional centrifugal force of the rotate Earth provided the vertical position and natural stability of tower.

The bottom part of tower located in troposphere has the bracing wires 4 which help the tower to resist the troposphere wind.

The control sectional conductivity layer allows to create the high voltage running wave which accelerates (and brakes) the cabins (as rotor of linear electrostatic engine) to any high speed. Electrostatic forces also do not allow the cabin to leave the tube.

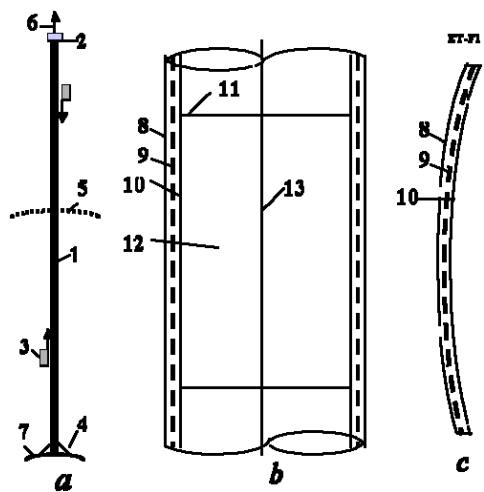

**Fig.6**. Electrostatic AB tower (mast, Space Elevator). (**a**) Side view, (**b**) Cross-section along axis, (**c**) Cross-section wall perpendicular axis. *Notations*: 1 - electrostatic AB tower (mast, Space Elevator); 2 - Top space station; 3 - passenger, load cabin with electrostatic linear engine; 4 - bracing (in troposphere); 5 - geosynchronous orbit; 6 - tensile force from electron gas; 7 - Earth; 8 - external layer of isolator; 9 - conducting control layer having sections; 10 - internal layer of isolator; 11 - internal dielectric partition; 12 - electron gas, 13 - laser control beam.

**2. Electron gas and AB tube.**   The electron gas consists of conventional electrons. In contract to molecular gas the electron gas has many surprising properties. For example, electron gas (having same mass density) can have the different pressure in the given volume. Its pressure depends from electric intensity, but electric intensity is different in different part of given volume. For example, in our tube the electron intensity is zero in center of cylindrical tube and maximum at near tube surface.

  The offered AB-tube is main innovation in the suggested tower. One has a positive control charges isolated thin film cover and electron gas inside. The positive cylinder create the zero electric field inside the tube and electron conduct oneself as conventional molecules that is equal mass density in any points. When kinetic energy of electron is less then energy of negative ionization of the dielectric cover or the material of the electric cover does not accept the negative ionization, the electrons are reflected from cover. In other case the internal cover layer is saturated by negative ions and begin also to reflect electrons. Impotent also that the offered AB electrostatic tube has neutral summary charge in outer space.

  **Advantages of electrostatic tower**. The offered electrostatic tower has very important advantages in comparison with space elevator:

1. Electrostatic AB tower (mast) may be built from Earth's surface without rockets. That decreases the cost of electrostatic mast in thousands times.
2. One can have any height and has a big control load capacity.
3. In particle, electrostatic tower can have the height of a geosynchronous orbit (37,000 km) WITHOUT the additional continue the space elevator (up 120,000 ÷ 160,000 km) and counterweight (equalizer) of hundreds tons.
4. The offered mast has less the total mass in tens of times then conventional space elevator.
5. The offered mast can be built from lesser strong material then space elevator cable (comprise the computation here and in [13] Ch.1).
6. The offered tower can have the high speed electrostatic climbers moved by high voltage electricity from Earth's surface.
7. The offered tower is more safety against meteorite then cable space elevator, because the small meteorite damaged the cable is crash for space elevator, but it is only create small hole in electrostatic tower. The electron escape may be compensated by electron injection.
8. The electrostatic mast can bend in need direction when we give the electric voltage in need parts of the mast.

    The electrostatic tower of height 100 ÷ 500 km may be built from current artificial fiber material in present time. The geosynchronous electrostatic tower needs in more strong material having a strong coefficient $K \geq 2$ (whiskers or nanotubes, see below).

**3. Other applications of offered AB tube idea**.

  The offered AB-tube with the positive charged cover and the electron gas inside may find the many applications in other technical fields. For example:

1) *Air dirigible*. (1) The airship from the thin film filled by an electron gas has 30% more lift force then conventional dirigible filled by helium. (2) Electron dirigible is significantly cheaper then same helium dirigible because the helium is very expensive gas. (3) One does not have problem with changing the lift force because no problem to add or to delete the electrons.
2) *Long arm*. The offered electron control tube can be used as long control work arm for taking the model of planet ground, rescue operation, repairing of other space ships and so on [13] Ch.9.
3) *Superconductive or closed to superconductive tubes*. The offered AB-tube must have a very low electric resistance for any temperature because the electrons into tube to not have ions and do not loss energy for impacts with ions. The impact the electron to electron does not change the total impulse (momentum) of couple electrons and electron flow. If this idea is proved in experiment, that will be big breakthrough in many fields of technology.

4) *Superreflectivity*. If free electrons located between two thin transparency plates, that may be superreflectivity mirror for widely specter of radiation. That is necessary in many important technical field as light engine, multy-reflect propulsion [13] Ch.12 and thermonuclear power [21] Ch.11.

The other application of electrostatic ideas is Electrostatic solar wind propulsion [13] Ch.13, Electrostatic utilization of asteroids for space flight [13] Ch.14, Electrostatic levitation on the Earth and artificial gravity for space ships and asteroids [13, Ch.15], Electrostatic solar sail [13] Ch.18, Electrostatic space radiator [13] Ch.19, Electrostatic AB ramjet space propulsion [20], etc.[21].

**Project.** As the example (not optimal design!) author of [19] takes three electrostatic towers having: the base (top) radius $r_0 = 10$ m; $K = 2$; heights $H = 100$ km, 36,000 km (GEO); and $H = 120,000$ km (that may be one tower having named values at given altitudes); electric intensity $E = 100$ MV/m and 150 MV/m. The results of estimation are presented in Table 2.

**Table 2.** The results of estimation main parameters of three AB towers (masts) having the base (top) radius $r_0 = 10$ m and strength coefficient $K = 2$ for two $E = 100$, 150 MV/m.

| Value | $E$ MV/m | $H$=100 km | $H$=36,000km | $H$=120,000km |
|---|---|---|---|---|
| Lower Radius , m | - | 10 | 100 | 25 |
| Useful lift force, ton | 100 | 700 | 5 | 100 |
| Useful lift force, ton | 150 | 1560 | 11 | 180 |
| Cover thickness, mm | 100 | $1\times10^{-2}$ | $1\times10^{-3}$ | $0.7\times10^{-2}$ |
| Cover thickness, mm | 150 | $1.1\times10^{-2}$ | $1.2\times10^{-3}$ | $1\times10^{-2}$ |
| Mass of cover, ton | 100 | 140 | $3\times10^{3}$ | $1\times10^{4}$ |
| Mass of cover, ton | 150 | 315 | $1\times10^{4}$ | $2\times10^{4}$ |
| Electric charge, C | 100 | $1.1\times10^{4}$ | $3\times10^{5}$ | $12\times10^{5}$ |
| Electric charge, C | 150 | $1.65\times10^{4}$ | $4.5\times10^{5}$ | $1.7\times10^{6}$ |

**Conclusion.** The offered inflatable electrostatic AB mast has gigantic advantages in comparison with conventional space elevator. Main of them is follows: electrostatic mast can be built any height without rockets, one needs material in tens times less them space elevator. That means the electrostatic mast will be in hundreds times cheaper then conventional space elevator. One can be built on the Earth's surface and their height can be increased as necessary. Their base is very small.

The main innovations in this project are the application of electron gas for filling tube at high altitude and a solution of a stability problem for tall (thin) inflatable mast by control structure.

## 6. Electromagnetic Space Towers (AB-Levitron) [20].

The AB-Levitron uses two large conductive rings with very high electric current (fig.7). They create intense magnetic fields. Directions of the electric currents are opposed one to the other and the rings are repelling, one from another. For obtaining enough force over a long distance, the electric current must be very strong. The current superconductive technology allows us to get very high-density electric current and enough artificial magnetic field at a great distance in space.

The superconductive ring does not spend net electric energy and can work for a long time period, but it requires an integral cooling system because current superconducting materials have a critical temperature of about 150-180 K. This is a *cryogenic* temperature.

However, the present computations of methods of heat defense (for example, by liquid nitrogen) are well developed and the induced expenses for such cooling are small.

The ring located in space does not need any conventional cooling—there, defense from Sun and Earth radiations is provided by high-reflectivity screens. However, a ring in space must have parts open to outer space for radiating of its heat and support the maintaining of low ambient temperature. For variable direction of radiation, the mechanical screen defense system may be complex. However, there are thin layers of liquid crystals that permit the automatic control of their energy reflectivity and transparency and the useful application of such liquid crystals making it easier for appropriate space cooling system. This effect is used by new man-made glasses that can grow dark in bright solar light.

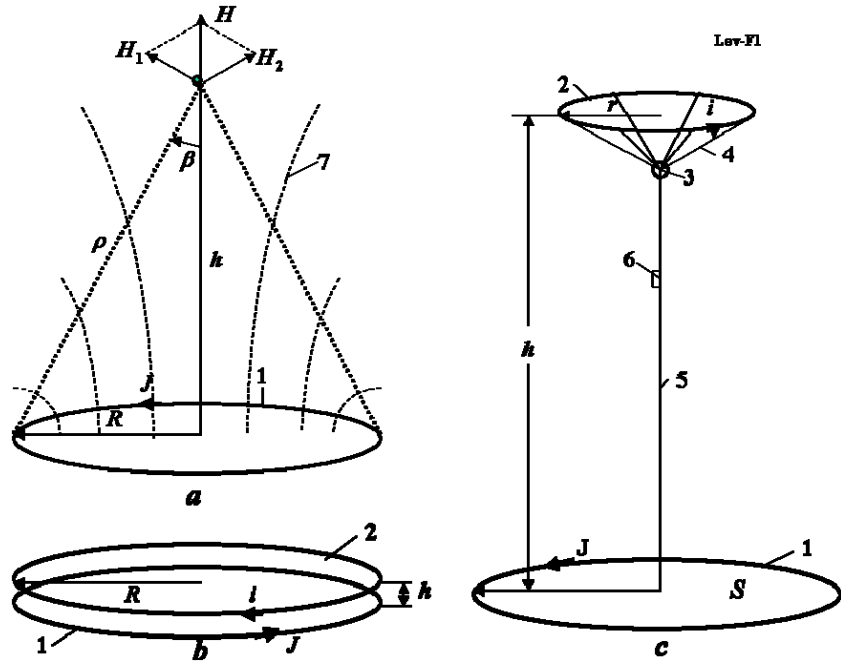

**Figure 7.** Explanation of AB-Levitron Tower. (**a**) Artificial magnetic field; (**b**) AB-Levitron from two same closed superconductivity rings; (**c**) AB-Levitron - motionless satellite, space station or communication mast. *Notation*: 1- ground superconductivity ring; 2 - levitating ring; 3 - suspended stationary satellite (space station, communication equipment, etc.); 4 - suspension cable; 5 - elevator (climber) and electric cable; 6 - elevator cabin; 7 - magnetic lines of ground ring; $R$ - radius of lover (ground) superconductivity ring; $r$ - radius of top ring; $h$ - altitude of top ring; $H$ - magnetic intensity; $S$ - ring area.

The most important problem of the AB-Levitron is the stability of the top ring. The top ring is in equilibrium, but it is out of balance when it is not parallel to the ground ring. Author offers to suspend a load (satellite, space station, equipment, etc) lower than this ring plate. In this case, a center of gravity is lower a net lift force and the system then become stable.

For mobile vehicles the AB-Levitron can have a running-wave of magnetic intensity which can move the vehicle (produce electric current), making it significantly mobile in the traveling medium.

**Project #1. Stationary space station at altitude 100 km.** The author of [20] estimates the stationary space station located at altitude $h = 100$ km. He takes the initial data: Electric current in the top superconductivity ring is $i = 10^6$ A; radius of the top ring is $r = 10$ km; electric current in the superconductivity ground ring is $J = 10^8$ A; density of electric current is $j = 10^6$ A/mm$^2$; specific mass of wire is $\gamma = 7000$ kg/m$^3$; specific mass of suspending cable and lift (elevator) cable is $\gamma = 1800$ kg/m$^3$; safety tensile stress suspending and lift cable is $\sigma = 1.5 \times 10^9$ N/m$^2 = 150$ kg /mm$^2$; $\alpha = 45^{\circ}$, safety superconductivity magnetic intensity is $B = 100$ T. Mass of lift (elevator) cabin is 1000 kg.

Then the optimal radius of the ground ring is $R = 81.6$ km (Eq, (3)[20], we can take $R = 65$ km); the mass of space station is $M_S = F = 40$ tons (Eq.(2)). The top ring wire mass is 440 kg or together with control screen film is $M_r = 600$ kg. Mass of two-cable elevator is 3600 kg; mass of suspending cable is less 9600 kg, mass of parachute is 2200 kg. As the result the useful mass of space station is $M_u$ = 40 - (0.6+1+3.6+9.6+2.2) = 23 tons.

Minimal wire radius of top ring is $R_T$ = 2 mm (Eq. (10)[20]). If we take it $R_T$ = 4 mm the magnetic pressure will be $P_T$ =100 kg/mm². Minimal wire radius of the ground ring is $R_T$ = 0.2 m. If we take it $R_T$ = 0.4 m the magnetic pressure will me $P_T$ =100 kg/mm². Minimal rotation speed (take into consideration the suspending cable) is $V$ = 645 m/s, time of one revolution is $t$ = 50 sec. Electric energy in the top ring is small, but in the ground ring is very high $E = 10^{14}$ J. That is energy of 2500 tons of liquid fuel (such as natural gas, methane).

The requisite power of the cooling system for ground ring is about $P$ = 30 kW.

**2**. **Magnetic Suspended AB-Structures** [22]. These structures use the special magnetic AB-columns [Fig. 8]. Author of [22] computed two projects: suspended moveless space station at altitude 100 km and the geosynchronous space station at altitude 37,000 km. He shows that space stations may be cheap launched by current technology (magnetic force without rockets) and climber can have a high speed.

As the reader observes, all parameters are accessible using existing and available technology. They are not optimal.

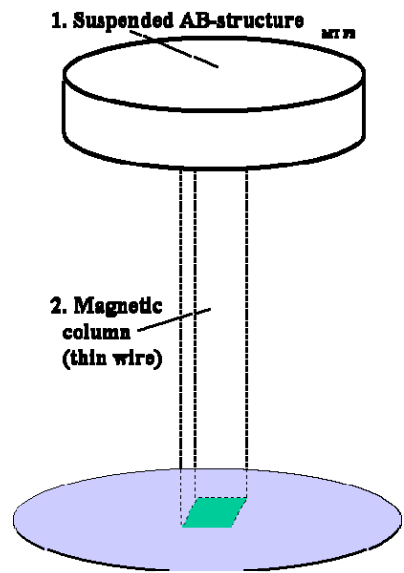

**Fig.8**. Suspended Magnetic AB-Structure

## General conclusion

Current technology can build the high height and space towers (mast). We can start an inflatable or steel tower having the height 3 km. This tower is very useful (profitable) for communication, tourism and military. The inflatable tower is significantly cheaper (in ten tines) then a steel tower, but it is having a lower life times (up 30-50 years) in comparison the steel tower having the life times 100 – 200 years. The new advance materials can change this ratio and will make very profitable the high height towers. The circle, kinetic, electrostatic and magnetic space towers promise a jump in building of space towers but they are needed in R&D.

search "Bolonkin")